\documentclass[a4paper]{jpconf}
\usepackage{citesort}
\usepackage{amsmath}
\usepackage{amsfonts}
\usepackage{amssymb}
\usepackage{graphicx}
\begin{document}
\title{Interplay of Local-Moment Ferromagnetism and Superconductivity in ErRh$_4$B$_4$ Single Crystals}

\author{R Prozorov, M D Vannette, S A Law, S L Bud'ko, P C Canfield}

\address{Ames Laboratory and Department of Physics \& Astronomy, Iowa State University, Ames, IA 50011, U.S.A}

\ead{prozorov@ameslab.gov}

\begin{abstract}
Tunnel-diode resonator technique was used to study crystals of ferromagnetic re-entrant superconductor, ErRh$_{4}$B$_{4}$. At the boundary between ferromagnetism (FM) and superconductivity (SC), dynamic magnetic susceptibility, $\chi(T,H)$, exhibits highly asymmetric behavior upon warming and cooling as well as enhanced diamagnetism on the SC side. SC phase nucleates upon warming in a cascade of discontinuous jumps in magnetic susceptibility $\chi(T,H)$, whereas FM phase develops gradually as reported in detail in \cite{prozorov2008}. Here we further investigate enhanced diamagnetism. We find that when a magnetic field is applied along the magnetic easy axes, a region of enhanced diamagnetic screening is smaller than in the perpendicular orientation. A discussion of possible causes of this effect is provided.
\newline\newline
\textit{Date: 15 June 2008}
\end{abstract}

Superconductivity (SC) and local-moment ferromagnetism (FM) coexisting in the same volume is an interesting theoretical and experimental problem \cite{Ginzburg1957,Bulaevskii1985,Sinha1989,Fischer1990,Maple1995,Kulic2006}. To this day, there are only few, confirmed, \textit{local-moment} ferromagnetic
superconductors, Ho$_{x}$Mo$_{6}$S$_{8}$ \cite{Ishikawa1977} ($T_{FM}\approx0.7$ K, $T_{c}\approx1.8$ K) and much studied ErRh$_{4}$B$_{4}$ \cite{Fertig1977,Machida1984,Bulaevskii1985,Sinha1989,Fischer1990,Maple1995,Kulic2006} ($T_{FM}\approx 1$ K, $T_{c}\approx 8.5$K). Here $T_{c}$ is a superconducting transition temperature and $T_{FM}$ is the ferromagnetic Curie temperature, although the latter is more difficult to determine, because FM develops on a SC background.

In ErRh$_{4}$B$_{4}$ superconductor the ferromagnetic phase is primitive tetragonal with spontaneous magnetization along the $a-$ axis. Anisotropies of the first, $H_{c1}$, and second, $H_{c2}$, critical fields were studied in detail by Crabtree \textit{et al.} \cite{Crabtree1982,Crabtree1986}. When a magnetic field was oriented along the easy magnetic $a-$ axis, $H_{c2}^{a}$ exhibited a peak at 5.5 K interpreted to be due to large paramagnetic spin susceptibility in that direction \cite{Crabtree1982,Fischer1990}. For a magnetic field applied along the hard $c-$ axis, $H_{c2}^{c}$ collapses near the onset of FM.

Much of prior work has focused on the question of the microscopic structure of the coexisting phase. Neutron diffraction found modulated ferromagnetic structure with a period of $\sim 10$ nm \cite{Moncton1980,Sinha1982}. Other measurements suggested patches of SC and FM phases as large as $\sim200$ nm in size with SC regions still containing modulated FM moment with a period of $\sim10$ nm \cite{Sinha1982}. Most reports noted a profound hysteresis of the measured properties, temperature asymmetric upon warming and cooling. This apparent first-order transition is consistent with the spiral state suggested by Blount and Varma \cite{Blount1979,Matsumoto1979}. Yet, other measurements found a continuos transition, for example in neutron diffraction \cite{Moncton1977,Moncton1980} and specific heat experiments \cite{DePuydt1986}. Such second-order transition can be realized in a modulated structure or via spontaneous vortex phase.

Other theoretical models of the coexisting regime include "cryptoferromagnetic" phase \cite{Anderson1959,Bulaevskii1985}, vortex lattice modulated spin structure \cite{Tachiki1979}, type-I superconductivity \cite{Tachiki1979,Gray1983,Bulaevskii1985}, gapless superconductivity and possibly, an inhomogeneous Fulde-Ferrell-Larkin-Ovchinnikov (FFLO) state \cite{Machida1984,Bulaevskii1985}. Another interesting possibility is the
development of superconductivity within the ferromagnetic domain walls
\cite{Buzdin1984,Buzdin2003}. Finally, Bulaevskii pointed out importance of the demagnetization factor in determining the nature of the $FM \leftrightarrow SC$ boundary \cite{Bulaevskii1985}. In our recent paper, $\chi(T,H)$, in the narrow temperature range of this coexisting region, was analyzed in detail \cite{prozorov2008}. The observed behavior was consistent with type-I like superconductor at the $SC/FM$ boundary. In this contribution we report anisotropic anomalous diamagnetic response close to a $FM$ phase that could be consistent with the development of an FFLO state.

The needle-shaped single crystals of ErRh$_{4}$B$_{4}$ were grown at high temperatures from a molten copper flux as described in \cite{Okada1996,Shishido1997}. The crystallographic c-axis was along the needles. $\chi$ was measured with a tunnel-diode resonator (TDR) which is sensitive to changes in susceptibility $\Delta\chi\sim10^{-8}$ \cite{Prozorov2006}.

\begin{figure}[t]
\includegraphics[width=9cm]{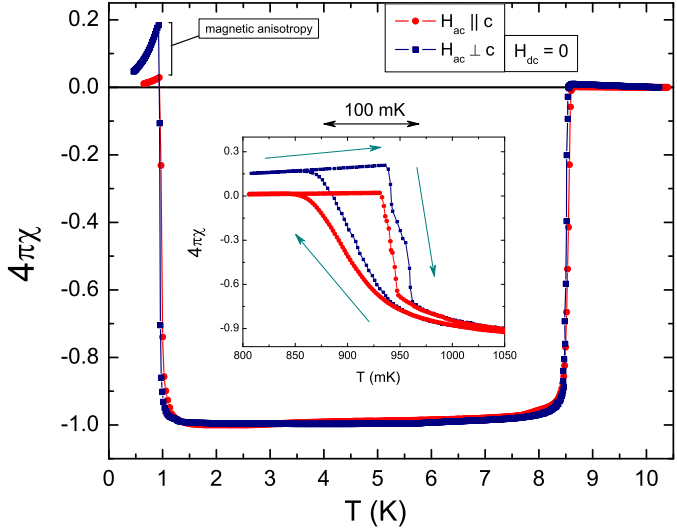}\hspace{2pc}%
\begin{minipage}[b]{6.1cm}\caption{\label{fig1}$4\pi\chi$ measured at \\ $H_{dc}=0$ for $H_{ac}$ applied in two orientations with respect to crystal axes. Red circles show \\$H \parallel c-axis$ and blue squares show $H \perp c-axis$ data. Note large magnetic anisotropy at the FM/SC boundary associated with easy/hard magnetic axes. Inset shows detail of the transition with directions of warming and cooling shown by arrows. The transition is only 0.1 mK wide, but shows highly asymmetric, jerky on warming, curve.}
\end{minipage}
\end{figure}

\begin{figure}[h]
\begin{center}
\includegraphics[width=15cm]{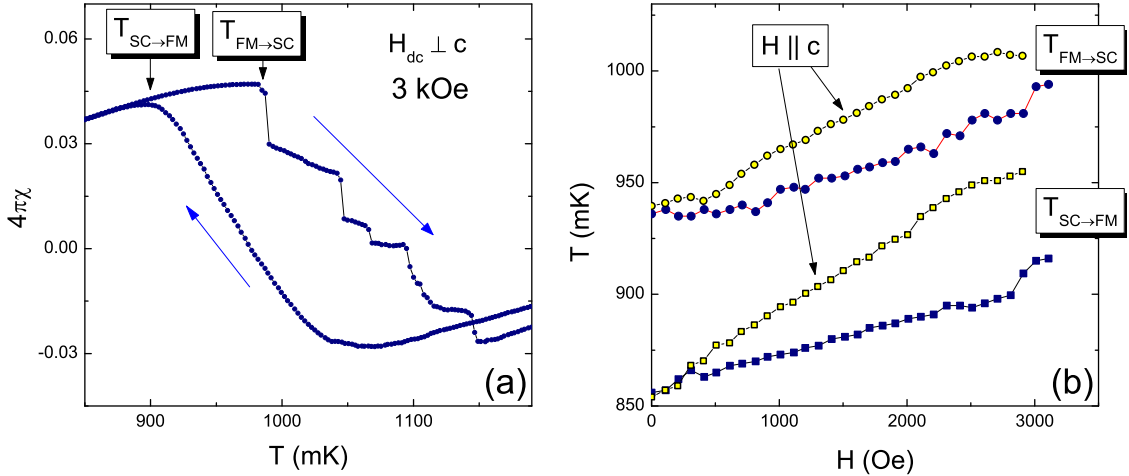}
\end{center}
\caption{\label{fig2}(a) $4 \pi \chi(T,H=3 kOe)$ in the transition region with graphical definitions of the onset, $T_{FM \rightarrow SC}$, and last signature of diamagnetic signal, $T_{SC \rightarrow FM}$, shown by arrows. (b) magnetic field dependence of the transition temperatures.}
\end{figure}

Figure \ref{fig2} shows magnetic field evolution of the transition from FM to SC state marked by $T_{FM \rightarrow SC}$ upon warming and by $T_{SC \rightarrow FM}$ upon cooling. Transition temperatures shift much more when a magnetic field is oriented perpendicular to the magnetic easy axes. Within a model proposed in our previous work, this behavior can be explained by noting that superconducting regions must be sandwiched between ferromagnetic domains with walls oriented perpendicular to the hard $c-$ axis. When an external field is applied along the $a-$ axis, it is only a small addition to the internal Weiss field of a ferromagnet, thus $T_{SC \leftrightarrow FM}$ is changed only moderately. However, when field is applied perpendicular to the domains, it tilts the magnetic moments inducing large fields in the regions where SC would have appeared. Therefore the change in $T_{SC \leftrightarrow FM}$ is much larger. It also implies that at least between  $T_{FM \rightarrow SC}$ and $T_{SC \rightarrow FM}$ the system is ferromagnetic.

\begin{figure}[t]
\begin{center}
\includegraphics[width=15cm]{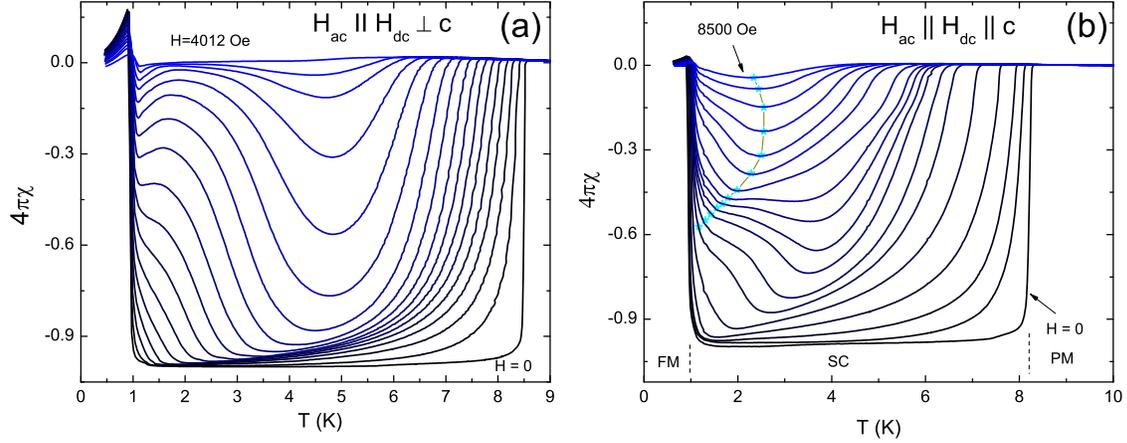}
\end{center}
\caption{\label{fig3}$4 \pi \chi(T)$ measured in two orientations at different magnetic fields. The anomalous screening is best seen in (b) is marked by a line.}
\end{figure}
\begin{figure}[b]
\begin{center}
\includegraphics[width=15cm]{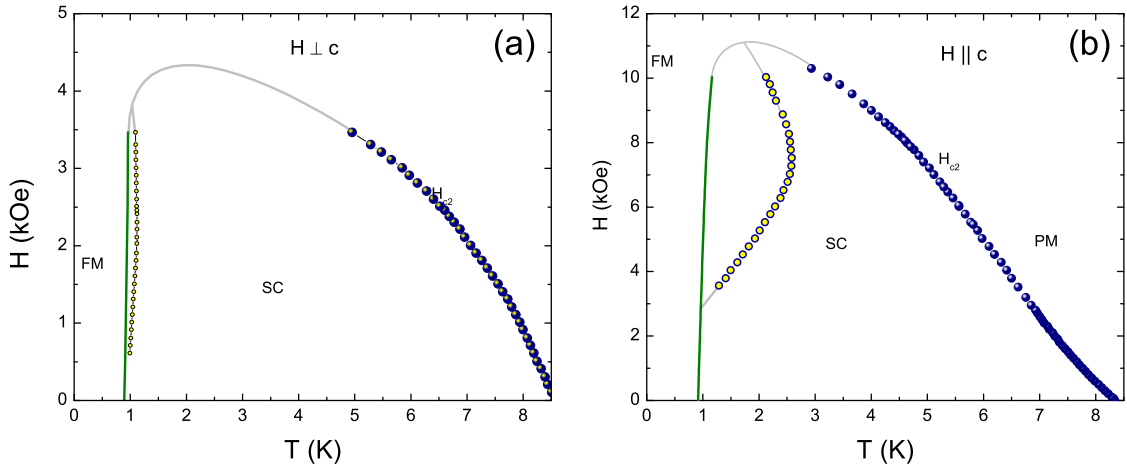}
\end{center}
\caption{\label{fig4}$H-T$ phase diagram of ErRh$_4$B$_4$ single crystal in two orientations. In addition to FM boundary and second critical field, temperature of the minimum in $4 \pi \chi (T)$} (maximum diamagnetism) is also shown.
\end{figure}

The difference between different orientations is clearly seen in Fig.~\ref{fig3}, which shows $4 \pi \chi(T)$ measured at different values of an applied magnetic field. An important novel feature is that diamagnetic response seem to be enhanced close to the FM boundary. Note that we do not attribute any significance to a dip at $\sim 4$ K, which is simply due to a crossover between two normal regions in a FM and PM states. Figure \ref{fig4} shows the phase diagram constructed from the measurements shown in Fig.~\ref{fig3}. These diagram resembles the one proposed by Bulaevskii 20 years ago, see page 197 in Ref.~\cite{Bulaevskii1985}. In that work, samples of different shape were predicted to demonstrate different behavior in the vicinity of the FM/SC border. In particular, our Fig.~\ref{fig4}(a) would map onto Fig.~7(a) of Ref.~\cite{Bulaevskii1985}. Although we could not change sample shape to explore other demagnetization factors, turning magnetic field in our interpretation is equivalent to an increase of demagnetization, as follows from our discussion of Fig.~\ref{fig2}. If so, we may expect development of an FFLO state in this orientation and Fig.~\ref{fig4}(b) maps well onto Fig.~7(c) of Ref.~\cite{Bulaevskii1985} that exhibits a large FFLO pocket. Indeed, this explanation is speculative without direct confirmation of the microscopic modulation of the order parameter. However, we conclude that our model \cite{prozorov2008} in which superconducting regions form between ferromagnetic domains remains plausible.

\ack
Discussions with Lev Bulaevskii, Alexander Buzdin, Vladimir Kogan, Kazushige Machida and Roman Mints are appreciated. Work at the Ames Laboratory was supported by the Department of Energy-Basic Energy Sciences under Contract No. DE-AC02-07CH11358. R. P. acknowledges support from the Alfred P. Sloan Foundation.

\section*{References}
\bibliography{ErRh4B4}
\end{document}